\documentclass[twoside,slac_one]{revtex4}
\usepackage{graphicx}
\usepackage{fancyhdr}
\usepackage{amsmath} 
\usepackage{bm}
\usepackage{amsxtra}
\usepackage{amssymb}
\usepackage{amsthm}
\usepackage{latexsym}
\usepackage{lscape}
\usepackage{xspace}    

\newcommand{\fluence}{\ensuremath{\textrm{n}_{\textrm{eq}}/\textrm{cm}^2}\xspace}
\newcommand{\ifb}{\ensuremath{\textrm{fb}^{-1}\xspace}}
\begin{document}

\title{Study of the readout chip and silicon sensor degradation for the CMS pixel upgrade}

%

\author{G. Tinti for the NSF PIRE group}
\affiliation{Department of Physics \& Astronomy, University of Kansas, Lawrence, KS, USA}

\begin{abstract}
Hybrid silicon pixel detectors are currently used in the innermost tracking system of the Compact Muon Solenoid (CMS) experiment. Radiation tolerance up to fluences expected for a few years of running of the Large Hadron Collider (LHC) has already been proved, although some degradation of the part of the silicon detector closer to the interaction point is expected. During the LHC upgrade phases, the level of dose foreseen for the silicon pixel detector will be much higher. To face this aspect, dedicated irradiation tests with fluences above $\mathcal{O}(10^{15})$~n$_{\textrm{eq}}$/cm$^2$ have been performed on the silicon sensor and readout chip. Changes in the operation of the sensor and readout chip as a function of the fluence are presented. The charge collection efficiency has been studied: partial recovery of the detector efficiency can be achieved by operating the detectors in a controlled environment and at higher bias voltage.
\end{abstract}

\maketitle

\thispagestyle{fancy}

\section{Introduction}
The LHC accelerator and the CMS experiment have performed very well in the last year, delivering an integrated luminosity of 3~\ifb up to the time of these proceedings (September 2011), with maximum instantaneous luminosity of 2.86~$\times 10^{33} \textrm{cm}^{-2} \textrm{s}^{-1}$. The silicon pixel modules are exposed to the highest ionization due to their proximity to the interaction point. A future upgrade of the LHC performance, foreseen in 2016, will push towards reaching high radiation fluences sooner, as the instantaneous luminosity after the upgrade will increase to $2 \times 10^{34} \textrm{cm}^{-2} \textrm{s}^{-1}$ and the radial distance of the first layer of the pixel detector will possibly be reduced. For these reasons, the limits of the radiation hardness for both the sensor and the readout chip is important to determine the detector lifetime.     


Section~\ref{CMSpixel} introduces the characteristics of the CMS pixel detector. Possible changes for the upgrade plan are also discussed. The effects that the radiation damage has on the sensor and the readout chip are presented in Section~\ref{limits}. The measurements of the performances of irradiated chips and sensors are presented in Section~\ref{measurements}. In particular, the variation of the collected charge induced by a minimum ionizing particle is presented. The performance of both the sensor and readout chips is examined. Section~\ref{risetime} is dedicated to the specific changes in the characteristic time of the rising edge of the pulse height signal. Conclusions are presented in Section~\ref{conclusions}.

\section{The CMS sensor and readout chip} \label{CMSpixel}
The CMS pixel detector~\cite{tdr,wolfram} are ``hybrid'' detectors as the silicon sensor is separated from the silicon readout chip. Each pixel, which has a size of $100~\mu \textrm{m} \times 150~\mu \textrm{m}$, has an indium bump bond connecting the sensor to the readout chip. The sensor has $n^+$ implants on the $n$ substrate and the $p$ junction on the backside. This choice allows the collection of electrons, which have a higher mobility and larger Lorentz angle (i.e., the charge sharing for tracks not perpendicular to the pixel front surface is enhanced). In addition, to extend the radiation hardness tolerance, the substrate is enriched with oxygen. Guard rings on the back side put the sensor edges at ground potential, limiting the risks of sparks at the edges. An extensive review of the radiation hardness property of $n^+$-in-$n$ sensors is available in~\cite{book}. 

A total of 52$\times$80 pixels are read by the PSI46V2 readout chip (ROC), which is described in~\cite{ROC}. For each single pixel, the signal from the sensor enters two charge sensitive stages: the preamplifier and the shaper. The zero suppression is performed by a comparator. Ideally, the minimum threshold should be as low as possible but above the noise and cross-talk levels. Lowering the threshold to the minimum improves the tracking reconstruction (as the charge sharing between pixels is exploited) and extends the lifetime of the sensor.  

The CMS pixel detector system is divided into a three layer barrel and two disk end-caps. The first layer of the barrel, which has a radial distance of 4.4~cm from the interaction point, is the one most exposed to radiation. Two additional layers are placed at radii 7.3~cm and 11.2~cm. The disks are placed at a longitudinal distance $z = \pm 34.5$~cm and $z = \pm 46.5$~cm. This geometry allows very good physics performance, with coverage up to $|\eta|<2.5$ and at least two pixel hits for any track inside the coverage. A detailed description of the CMS pixel detector system and its readout is available in~\cite{wolfram}.  
 
The pixel detector in CMS is performing well~\cite{performances}. A Minimum Ionizing Particle (MIP) is expected to create about 22000 electrons. The average noise measured in the present CMS pixel detector is 120 electrons, much lower than the expected signal. However, the minimum threshold for a particle to be detected is on average 2500 electrons. The threshold, which is much higher than the noise, is due to the internal cross talk of the readout chip. Ongoing efforts are devoted to the understanding of different sources of internal cross talk and will be implemented in the new version of the PSI46 ROC.

An upgraded pixel detector is foreseen for the so-called ``phase~I upgrade'' in 2016. A new pixel detector with four barrel layers and three endcap disks is being designed. The new geometry minimizes the material budget while increasing the tracking points up to four. The high rates foreseen justify the study of a fast digital readout architecture to ensure readout efficiency. From the sensor point of view, the major change, in addition to the increase of the rate of events, is the new bi-phase CO$_2$ cooling system~\cite{CO2}, which will allow pixel operation at $-20^\circ$C coolant temperature. More details of the upgrade project are given in these proceedings~\cite{dpfupgrade}. The present $\textrm{C}_6 \textrm{F}_{14}$ cooling system has also been designed to run at $-20^\circ$C coolant temperature, but problems in operating the pixel detector due to the high humidity require to run at $+7^\circ$~C coolant temperature. Investigations are ongoing to reach a lower operating temperature keeping the humidity under control. 

\section{Radiation damage}\label{limits}   
The LHC accelerator has already delivered nearly 3~\ifb. For the first layer of the barrel pixel detector, 1~\ifb corresponds to a measured fluence of $2.4 \times 10^{12}$~\fluence. Up to 2016, when the phase~I detector is installed, the predicted integrated luminosity is 70~\ifb, or approximately a fluence of $2 \times 10^{14}$~\fluence at the innermost layer. In an extreme scenario where the current pixel detector will be functioning up to an integrated luminosity of 250~\ifb, the fluence is estimated to be $\mathcal{O}(10^{15})$~\fluence.

The effects of the radiation damage on the silicon sensor have been extensively described in~\cite{book}. We can expect three major effects:
\begin{itemize}
\item \textbf{Increase in the leakage current of the sensor}: defects are created in the sensor and cause the leakage current to increase. The leakage current increase can be monitored to understand the radiation damage from the detector itself. At the integrated luminosity of 1~\ifb, the leakage current per pixel has been measured to be 0.75~nA at the innermost layer. The preamplifier is guaranteed to work up to at least 100~nA/pixel. The leakage current increase will cause a reduction of the pulse height estimated to be $\frac{-0.2\%}{\textrm{nA}}$ per pixel.  
\item \textbf{Charge trapping and partial depletion of the sensor}: there can be a loss in pulse height due to partial depletion or charge trapped in the crystal bulk before collection. The CMS pixel detector does not manifest signs of charge trapping and partial depletion at the present fluence. Partial recovery of the pulse height loss is achievable by increasing the bias voltage. 
\item \textbf{Readout chip damage:} the internal voltages can changed with radiation and affect the operation of the ROC. These kinds of effects are not observed yet for the CMS pixel detector.
\end{itemize} 

The studies presented in this paper intend to determine at which fluence the above radiation induced effects become visible and can affect the full operation of the detectors.

\section{Radiation hardness measurements}\label{measurements}

Samples identical to the present CMS silicon sensors and ROCs have been tested to determine the charge collected as a function of the fluence. The samples had been irradiated with protons of 26~MeV momentum at the Karlsruhe Institute of Technology proton cyclotron facility~\cite{kit} in 2010. In addition to unirradiated samples for reference, the studied samples have fluences of $3 \times 10^{14}$~\fluence, $6 \times 10^{14}$~\fluence and $1.2 \times 10^{15}$~\fluence. A sample of fluence $3 \times 10^{15}$~\fluence was also tested for the measurement presented in Section~\ref{risetime}. To avoid annealing, the samples were kept in a freezer at $-18^\circ$C when not under test. Preliminary current/voltage curves, which are shown in Figure~\ref{fig:iv}, confirm the goodness of the samples before testing. When under test, the samples were cooled through a Peltier cooler to $-20^\circ$C or $-25^\circ$C depending on the fluence. A stable temperature within $0.3^\circ$C was maintained along with a minimal humidity of less than 30\% during the measurement. The setup and procedures implemented here closely follow the measurements presented in~\cite{Rohe:2009,Rohe:2010}.

\begin{figure}[h]
\centering
\includegraphics[width=90mm]{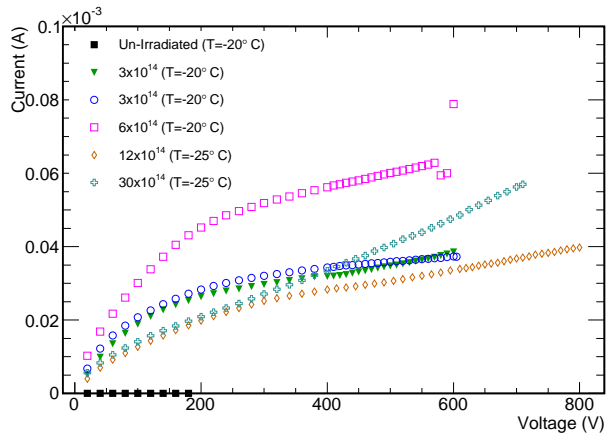}
\caption{Current versus bias voltage plots for the tested samples. The curves do not show any sign of breakdown or anomalous behavior of the tested detectors. The fluence and the temperature at which the samples were tested are reported in the legend. Note that the bias voltage applied and current are negative, but absolute values are graphed.} 
\label{fig:iv}
\end{figure}

The response of the detectors was studied using MIPs produced by the decay of $^{90}$Sr, which has a product endpoint of 2.28~MeV. The trigger was provided by a scintillator positioned underneath the pixel detector. Particles from the radioactive source had to travel both through the pixels and the scintillator to fire the trigger. In addition to MIPs, lower energy beta particles are also produced by decay of $^{90}$Sr. Those particles suffer from multiple scattering and give rise to multiple hit clusters in the pixels. To limit the effect of the low energy particles, only single hit clusters are considered in the analysis. This preferentially selects particles that are traveling from the source to the scintillator perpendicular to the pixels. Due to the uncertainty introduced by the multiple scattering, no absolute sensor efficiencies are reported in this paper.  

The most probable value of the charge deposited in the pixels has been determined by fitting the charge collected spectrum with a Landau function convoluted with a Gaussian. The fit has been performed by fitting only the peak region, in order to remove the bias introduced by the tail of the distribution. Figure~\ref{fig:landau} shows an example of a fitted spectrum.

\begin{figure}[h]
\centering
\includegraphics[width=90mm]{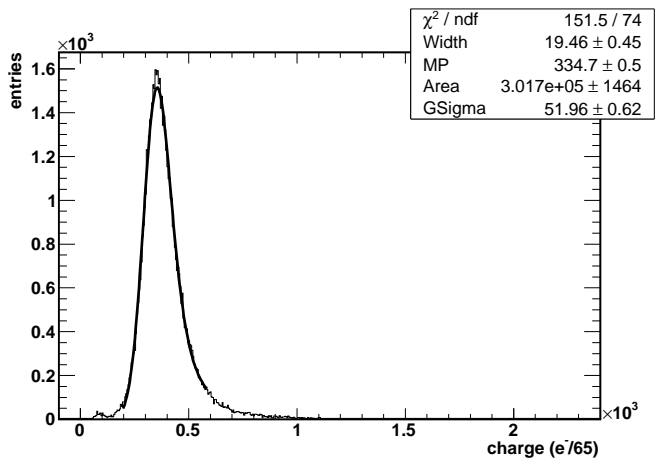}
\caption{Example of the distribution for the collected charge for a non irradiated sample at full depletion bias voltage. The most probable value of deposited charge is obtained by fitting the spectrum in the peak region with a Landau convoluted with a Gaussian, as shown in the plot.} 
\label{fig:landau}
\end{figure}

The collected charge has been studied as a function of the bias voltage. The bias voltage is actually negative, but absolute values are reported here. As it is clearly seen in Figure~\ref{fig: charge_bias}, by increasing the bias voltage the full depletion of the silicon substrate is achieved and the full charge is collected. In the case of the unirradiated samples, the plateau region is achieved already at a bias voltage of 100~V. When the fluence of the sample increases, it is still possible to achieve the full depletion by increasing the bias voltage up to at least 600~V for a fluence of $1.2 \times 10^{15}$~\fluence. The figure also shows that the pixel detector could be safely operated even at bias voltages as high as 1000~V.  

\begin{figure}[h]
\centering
\includegraphics[width=90mm]{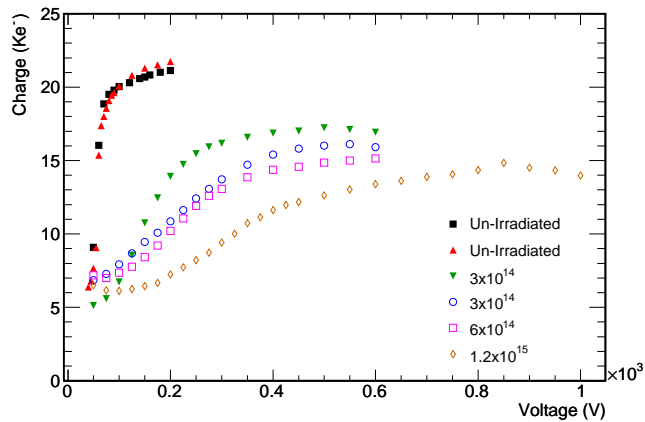}
\caption{Most probable charge versus bias voltage for the samples at different fluences. The plateau region, which indicates that the full depletion has been achieved, is visible at all fluences, but at higher bias voltages for irradiated samples.}
\label{fig: charge_bias}
\end{figure}
 
Figure~\ref{fig:charge_fluence} shows the charge collected as a function of the fluence once the full depletion is achieved. Even at a fluence of $1.2 \times 10^{15}$~\fluence, more than 13000 electrons are collected for a MIP. Even though the sensor efficiency is high even at high irradiation, the Lorentz angle, which enhances the charge sharing, diminishes at higher voltages. Separate studies from the one presented here are needed to determine the impact that the lowering of the Lorentz angle has on the spatial resolution.

\begin{figure}[h]
\centering
\includegraphics[width=90mm]{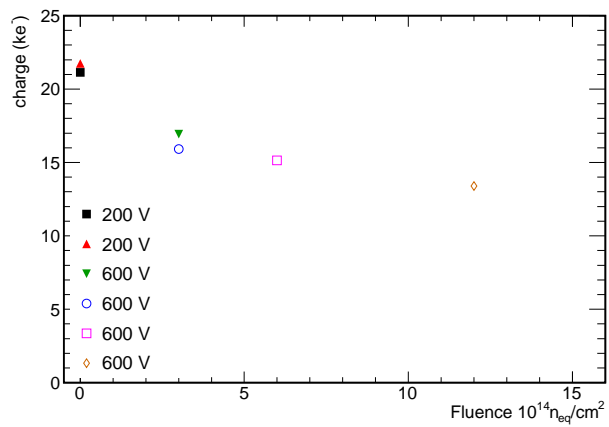}
\caption{Charge at full detector depletion as a function of the fluence. For unirradiated samples the collected charge is approximately 22000 electrons. At a fluence of $1.2 \times 10^{15}$~\fluence, the collected charge is about 13000 electrons. The measurements presented here agree with the ones reported in~\cite{Rohe:2010}.}   
\label{fig:charge_fluence}
\end{figure}

The readout chips were functioning with minimal changes to the nominal settings. Only small changes to the preamplifier and shaper feedback circuits were applied. The measured noise level has been stable even at high irradiation (at full depletion). In the measurements the threshold was set to 3900 electrons, which is low enough compared to the charge collected. For the sample at $3 \times 10^{15}$~\fluence, a full collected charge versus bias voltage scan had been collected, but the threshold of 3900 electrons was too high compared to the collected charge. A new collected charge versus bias voltage scan was planned with lower threshold, but the sample stopped functioning for unknown reasons.  

\section{Characteristic time measurement}\label{risetime}

The rise time of the amplifiers in the readout chip might change due to radiation. As the clock cycle of the ROC is 25~ns (equivalent to the nominal LHC bunch crossing interval), it is important that the pulse height is collected in time. Radiation damage could cause the preamplifier risetime to be slower.
The analog pulse shape curve of the PSI46V2 cannot be measured directly. An indirect measurement of the rising edge of the pulse shape can be achieved by scanning the delay between the calibration signal and the leading edge of the clock cycle that is readout. This method has already been explained in~\cite{Rohe:2010}. Here, a quantitative measurement of the characteristic time of the rising edge of the pulse height is performed. The pulse height (PH) behavior as a function of time (t) is parametrized as:        
\begin{equation}
PH(t)= p_0+ p_1 \cdot t \cdot e^{-\frac{t}{p_2}}
\label{eq:risetime}
\end{equation}
and fitted to the rising edge curve obtained from the measurement to obtain the parameters $p_0$, $p_1$ and $p_2$. Figure~\ref{fig:risetimefit} shows the pulse height rising edge and the corresponding fit. To balance the time required for the measurement and enough statistic to cancel the pixel by pixel variation, the measurements of the rising edge presented here are on a uniform sampling of the 25\% of the pixels of the ROC. 
 
\begin{figure}[h]
\centering
\includegraphics[width=90mm]{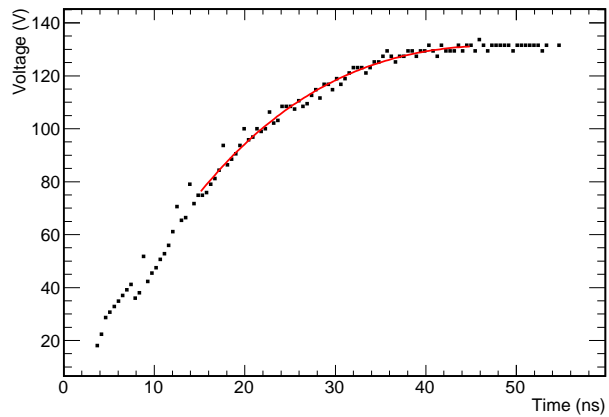}
\caption{Rising edge of the amplifier as obtained by an indirect scan, as explained in~\cite{Rohe:2010}. The curve obtained by fitting the rising edge with the function in equation~\eqref{eq:risetime} is superimposed in red.}  
\label{fig:risetimefit}
\end{figure}

The characteristic time is evaluated as the time at which the pulse height reaches 90\% of the value at a time $t=p_2$. For unirradiated ROCs without sensor, the characteristic time has been measured to be $(27 \pm 1)$~ns. For unirradiated ROCs with sensor, the measured time is $(30 \pm 1)$~ns. The sensor adds an extra capacitance in series to the readout chip. 

The measurement of the characteristic time has also been performed on the irradiated samples to record the variation as a function of the fluence. Table~\ref{tabtime} shows the measured characteristic times \footnote{The measurement at a fluence of $3 \times 10^{15}$~\fluence has actually been performed only on 10\% of the readout chip pixels. This could explain the slightly lower value obtained for this fluence compared to the other fluences.}.  

\begin{table}[h]
\caption{Characteristic amplifier risetimes as measured as a function of the fluence received by the sensors and ROCs under test.}
\begin{tabular}{|c|c|} 
\hline
Radiation ($10^{14}$~\fluence) & time (ns)\\
\hline
0  &      30 $\pm$ 1\\
\hline
3  &      38 $\pm$ 2\\
\hline
6  &      39 $\pm$ 3\\
\hline
12 &       39 $\pm$ 3\\
\hline
30 &       36 $\pm$ 2\\
\hline
\end{tabular}
\label{tabtime}
\end{table}
Exposing the sample to radiation increases the characteristic time of the preamplifier, but there is not a large difference depending on the irradiation received. This effect might suggest a saturation effect most likely from the sensor after irradiation. Specific tests need to be performed on irradiated bare ROCs (without the sensor) in order to decouple the effects. 

The preamplifier risetime, thus the characteristic time, depends on the choice of the analog current in the readout chip. In all the measurements presented here, the analog current has been fixed to 24~mA, using the nominal settings of the PSI46V2 ROC. An increase of the analog current should recover the irradiated risetime values to meet the value before irradiation, but specific studies have not been done yet. 

\section{Conclusions}\label{conclusions}
The present CMS sensor and readout chips have been tested up to fluences of $1.2 \times 10^{15}$~\fluence. By increasing the bias voltage to at least 600~V, the charge collected is more than 13000 electrons, ensuring good sensor efficiency and ROC behavior. The readout chip does not manifest major changes up to the tested fluences. A fluence of $1.2 \times 10^{15}$~\fluence corresponds to an integrated luminosity higher than 250~\ifb on the first layer of the barrel pixel detector and assures that the CMS pixel detector can function through the planned phase~I upgrade and even longer in case there is a delay. As the CMS pixel are running at somewhat warm temperatures due to humidity problems, the leakage current could become critical for the first layer. After the upgrade, the pixel detector will not suffer from leakage current so much as the operating temperature will be lowered to $-20^\circ$C~\cite{upgrade}. Further measurements are foreseen by the PSI and NSF PIRE groups in order to test ROCs up to a fluence of $5\times 10^{15}$~\fluence and to properly characterize the behavior of the amplifier risetime as a function of the fluence.

\begin{acknowledgments}
The work presented here could not be possible without the help of the PSI pixel group, in particular Dr. T.~Rohe, Dr. W.~Erdmann, Dr. H.-C.~K\"astli and J.~Sibille. Prof. A.~Bean (KU) has been precious to keep the measurement on track. The measurements were taken at PSI by the NSF PIRE group students, in particular by Joaquin Siado (UPRM). This work is supported by the PIRE grant OISE-0730173 of the US-NFS. 
\end{acknowledgments}



\end{document}